\documentclass[prl,twocolumn,showpacs,superscriptaddress]{revtex4-1}
\usepackage{graphicx}
\usepackage{bm}
\begin{document}


\title{Magnetism in superconducting EuFe$_2$As$_{1.4}$P$_{0.6}$ single crystals studied by local probes\/}
     \author{J. Munevar}
     \author{H. Micklitz}
     \author{M. Alzamora}
     \affiliation{Centro Brasileiro de Pesquisas Fisicas, Rua Xavier Sigaud 150, Rio de Janeiro, Brazil}
     \author{C. Arg\"uello}
     \affiliation{Department of Physics, Columbia University, New York, New York 10027, USA}
     \author{T. Goko}
     \affiliation{Department of Physics, Columbia University, New York, New York 10027, USA}
     \affiliation{TRIUMF, 4004 Wesbrook Mall, Vancouver, B.C., V6T 2A3, Canada}
     \author{F.~L.~Ning}
     \affiliation{Department of Physics, Columbia University, New York, New York 10027, USA}
      \author{A. A. Aczel}
     \affiliation{Department of Physics and Astronomy, McMaster University, Hamilton, Ontario L8S 4M1, Canada}
     \affiliation{Neutron Scattering Science Division, Oak Ridge National Laboratory, Oak Ridge, TN 37831, USA}
     \author{T. Munsie}
     \affiliation{Department of Physics and Astronomy, McMaster University, Hamilton, Ontario L8S 4M1, Canada}
     \author{T. J. Williams}
     \affiliation{Department of Physics and Astronomy, McMaster University, Hamilton, Ontario L8S 4M1, Canada}
      \author{G. F. Chen}
     \affiliation{Renmin University of China, Beijing 100872, China}
     \author{W. Yu}
     \affiliation{Renmin University of China, Beijing 100872, China}
   \author{G. M. Luke}
     \affiliation{Department of Physics and Astronomy, McMaster University, Hamilton, Ontario L8S 4M1, Canada}
     \affiliation{Canadian Institute of Advanced Research, Toronto, Ontario M5G 1Z8, Canada}
     \author{Y. J. Uemura}
     \affiliation{Department of Physics, Columbia University, New York, New York 10027, USA}
     \author{E. Baggio-Saitovitch}
     \email[author to whom correspondences should be addressed: E-mail ]{elisa@cbpf.br}
     \affiliation{Centro Brasileiro de Pesquisas Fisicas, Rua Xavier Sigaud 150, Rio de Janeiro, Brazil}
     \date{\today}
   \begin{abstract}
\noindent
{We have studied the magnetism in superconducting single crystals of EuFe$_2$As$_{1.4}$P$_{0.6}$ by using the local probe techniques of zero-field muon spin rotation/relaxation and $^{151}$Eu/$^{57}$Fe M\"ossbauer spectroscopy.  All of these measurements reveal magnetic hyperfine fields below the magnetic ordering temperature $T_M=$ 18 K of the Eu$^{2+}$ moments.  The analysis of the data shows that there is a coexistence of ferromagnetism, resulting from Eu$^{2+}$ moments ordered along the crystallographic c-axis, and superconductivity below $T_{SC}\approx$ 15 K.  We find indications for a change in the dynamics of the small Fe magnetic moments ($\sim$ 0.07 $\mu_B$) at the onset of superconductivity: below $T_{SC}$ the Fe magnetic moments seem to be ``frozen'' within the ab-plane.\/}
\end{abstract}
\maketitle

Since the initial observation of superconductivity in the  iron pnictides in 2008 \cite{hosono}, research in this field has been very intense, leading to the discovery of several Fe-containing compounds with superconducting transition temperatures $T_{SC}$ as high as 55 K in SmFeAsO$_{1-x}$F$_x$ \cite{smofeas}.  These FeAs-based compounds have a FeAs layer separated by a charge reservoir composed of RO (R is a rare earth, so-called 1111), alkali atoms or Eu(122), Li and Na atoms (111), or perovskite layers (42622).  This led to the classification of the Fe-containing superconductors in several families, with different properties among them mainly due to their various structures \cite{lumsden}.  The 122 family is particularly interesting because the structural and magnetic spin density wave (SDW) phase transition for the parent compound are at the same temperature, differing from the 1111 family where $T_S$  is above $T_N$ \cite{luetkens,goko,jiunhawchu,junzhao}.  

For the EuFe$_2$As$_2$ parent compound, the Fe lattice develops an antiferromagnetic SDW transition around 190 K, while Eu moments order antiferromagnetically around 20 K \cite{xiao}.  Superconductivity is induced by doping and external pressure, and the interplay between magnetic order and superconductivity has been extensively studied.  For example, superconductivity appears under external pressure at 30 K \cite{terashima}, and a reentrant behavior is found close to 20 K, where Eu moments align antiferromagnetically \cite{miclea}.  The effect of an external magnetic field is also remarkable, since a spin reorientation for Eu moments to ferromagnetic ordering is seen for an applied field as high as 3 T\cite{jiang1}.  Although doping on either the Eu or Fe sites also induces superconductivity \cite{jiang2,anupam}, the most interesting case may be doping the As site by P, where recent studies have implied a change of valence state for Eu \cite{lilingsun}, an increase of $T_{SC}$ up to 48.3 K under external pressure \cite{jingguo}, and evidence for coexistence of ferromagnetism and superconductivity\cite{jiang2,anupam,zhiren,1011.4481,ahmed}.  

Most recently, $^{151}$Eu and $^{57}$Fe M\"ossbauer studies on polycrystalline EuFe$_2$As$_{2-x}$P$_x$ samples were reported \cite{nowik}.  The results of those studies will be compared in detail with those of our paper.  We will show that due to the use of (i) single crystals and (ii) an additional local probe, namely muon spin rotation/relaxation ($\mu$SR), we are able to gain new insight on the interplay between magnetism and superconductivity in this compound.

In this work, we performed resistivity, magnetization and spectroscopic measurements ($^{57}$Fe/$^{151}$Eu M\"ossbauer and $\mu$SR) on EuFe$_2$As$_{1.4}$P$_{0.6}$ single crystals prepared by G.F. Chen's group in China. Resistivity  measurements were performed with a PPMS system using the transport mode at CBPF in Brazil, and magnetization was measured with a SQUID magnetometer by the McMaster group in Canada.  Zero field $\mu$SR spectra were taken at the M20 beamline in TRIUMF, Canada.  $^{57}$Fe and $^{151}$Eu Mossbauer studies were performed at CBPF in a $^4$He variable temperature cryostat,  moving the $^{57}$Co:Rh and $^{151}$Sm$_2$O$_3$ sources in sinusoidal mode and kept at the same temperature as the sample.  As an absorber, a single crystal mosaic was prepared from thin platelets ($t\approx 15$ $\mu m$), with the single crystal mosaic c-axis perpendicular to the absorber plane. 

\begin{figure}[t]

\begin{center}
\includegraphics[angle=0,width=8.0cm]{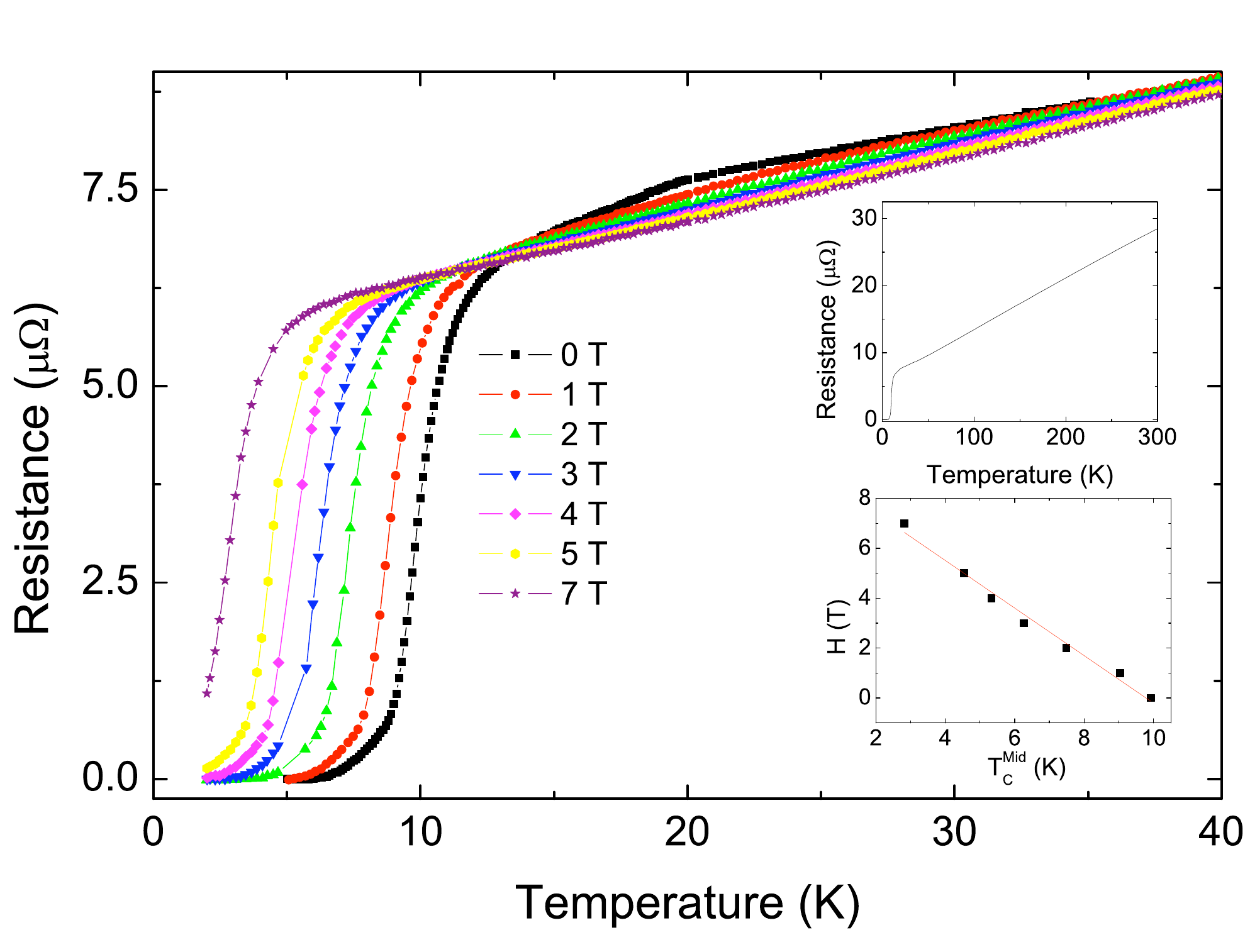} 
\caption{\label{Figure 1.} Four probe resistivity measurements for a small EuFe$_2$As$_{1.4}$P$_{0.6}$ single crystal, for external fields up to 7 T.  The onset and midpoint of the transition at 0 T are estimated to be 15 K and 10.5 K respectively. The two insets show the resistivity from 2 to 300 K without field, and $T_{mid}$ as a function of applied field.}
\end{center}
\end{figure}

Resistivity measurements were carried out on several crystals without external magnetic field, yielding for all of them the same drop in resistivity around 15 K, as shown in Fig.~\ref{Figure 1.}.  The metallic behavior of the single crystal above the transition, as expected for the iron pnictide superconductors, can be seen in the left inset of Fig.~\ref{Figure 1.}.  Application of an external magnetic field up to 7 T parallel to the c-axis results in a downshift of the resisitivity curves, giving $H_{c2}\approx10$ T.  A small deviation from linearity around 20 K is observed for the resistivity curve. It is suppressed as the external magnetic field increases and disappears for 7 T.  This behavior is probably caused by Eu spin fluctuations around the magnetic order at $T_M=$ 18 K, which are suppressed by the external field.  

\begin{figure}[t]

\begin{center}
\includegraphics[angle=0,width=8.0cm]{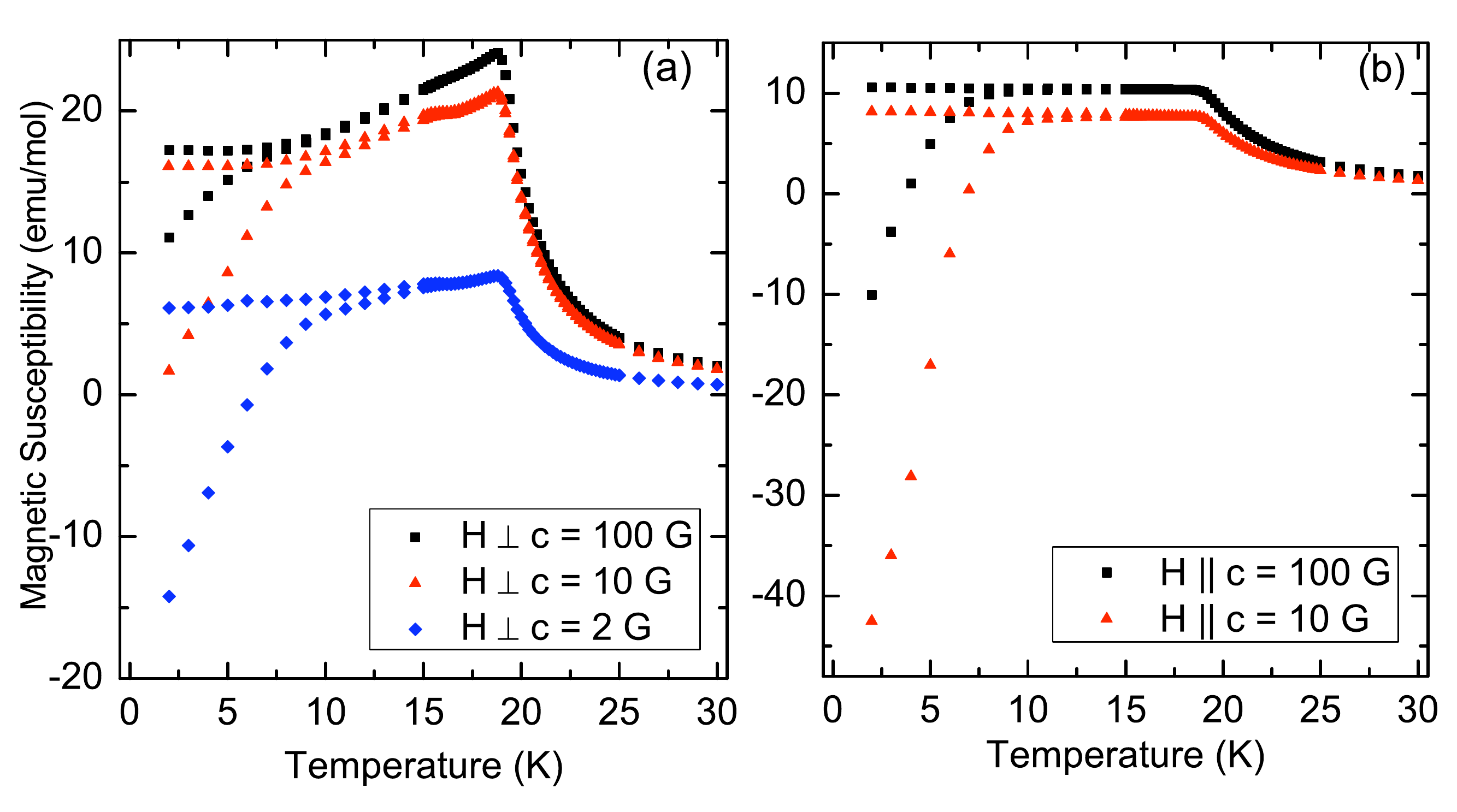} 
\caption{\label{Figure 2.} Magnetic susceptibility measurements performed for EuFe$_2$As$_{1.4}$P$_{0.6}$ single crystal, with the c-axis both (a) perpendicular and (b) parallel to the applied field.  A clear magnetic transition is observed at 18 K, followed by the superconducting transition around 15 K as evidenced by the Meissner effect.}
\end{center}
\end{figure}

The magnetic susceptibility of a EuFe2As$_{1.4}$P$_{0.6}$ single crystal was measured with a SQUID magnetometer with the c-axis both parallel and perpendicular to the applied magnetic field. The data shows an anomaly around 18 K that is related to the Eu moments ordering, and the Meissner effect is clearly observed below 15 K.  This inferred superconducting transition temperature agrees well with our resistivity results, and so we have both magnetism and superconductivity coexisting in the same sample. The magnetic order of the Eu$^{2+}$ sublattice is suspected to be ferromagnetic, as fitting our high temperature data in the paramagnetic regime to a Curie-Weiss law yielded $\theta=25$ K.  No further ordering coming from impurities or the Fe lattice is observed, so we assume that the doping has fully suppressed the SDW ordering for Fe atoms.

\begin{figure}[t]
\begin{center}
\includegraphics[scale=0.45]{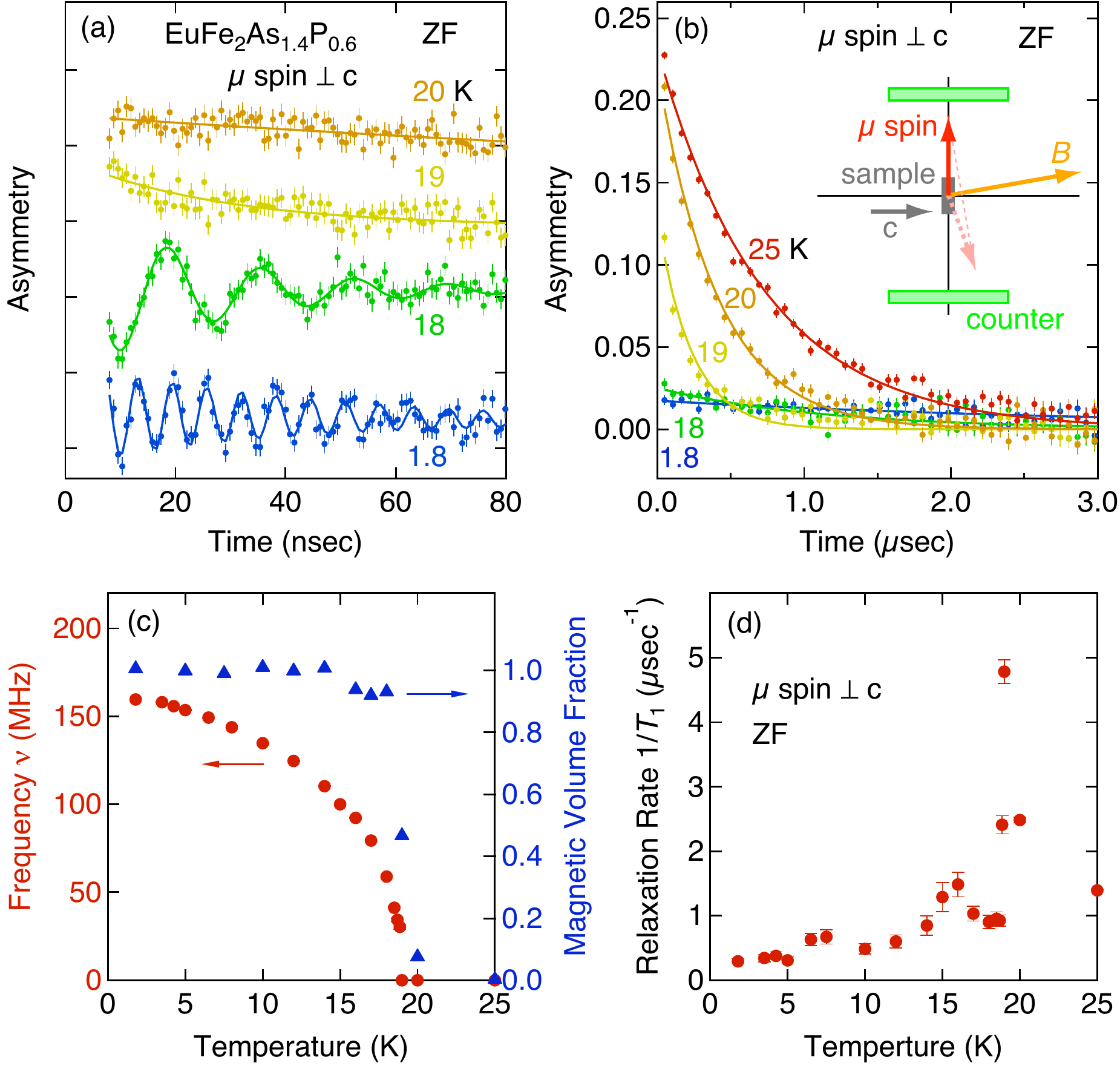} 
\caption{(Color online)
ZF-$\mu$SR time spectra with a time range of (a) 0-80 nsec 
and (b) 0-3 $\mu$sec
which were measured with the initial muon spin perpendicular to the c axis.
The spectra in (a) are vertically offset for clarify.
The inset of (b) displays a schematic overview of experimental setup
for the $\mu$ spin $\perp$ c orientation.
(c) Muon precession frequency and magnetic volume fraction 
as a function of temperature.
(d) Temperature dependence of
relaxation rate $1/T_{1}$ extracted from the ZF-$\mu$SR spectra in (b).
}
\end{center}
\end{figure}

Zero field $\mu$SR measurements were performed with the initial muon spin polarization perpendicular and parallel to the c axis of the crystal. A clear precession signal was observed in the ZF-$\mu$SR time spectra below $T_{\rm M}$ for $\mu$ spin $\perp$ c, as shown in Fig. 3(a). On the other hand,  oscillations were not observed for $\mu$ spin $\parallel$ c. These results indicate that the internal field $\bm{B}$ at the muon site is essentially parallel to the c axis. The muon precession frequency (Fig. 3(c))
corresponds to internal fields greater than 1 T at low temperature, indicating that the Eu$^{2+}$ ordered moments are quite large, as generally expected for this magnetic atom. The magnetic volume fraction 
(Fig.~3c) obtained  from the signal amplitudes in the ZF-$\mu$SR spectra for the two different initial muon spin orientations  indicates that Eu$^{2+}$ moments are ordered in almost the entire volume. Figure 3(d) shows temperature dependence of the muon relaxation rate $1/T_{1}$ which was derived from the spectra shown in Fig. 3(b). The relaxation rate exhibits divergent behavior at $T_{\rm M}$, suggesting the critical slowing down of Eu$^{2+}$ moments. There is a small peak of $1/T_{1}$ at $T \sim$ 16 K, which might be related to some magnetic anomaly at the onset of the superconducting state. We attempted  to detect the superconducting response of the single crystals with $\mu$SR by searching for flux expulsion or the field-broadening effects of a vortex lattice, but it was not possible due to the intrinsic magnetism of the sample.

\begin{figure}[t]

\begin{center}
\includegraphics[angle=0,width=8.0cm]{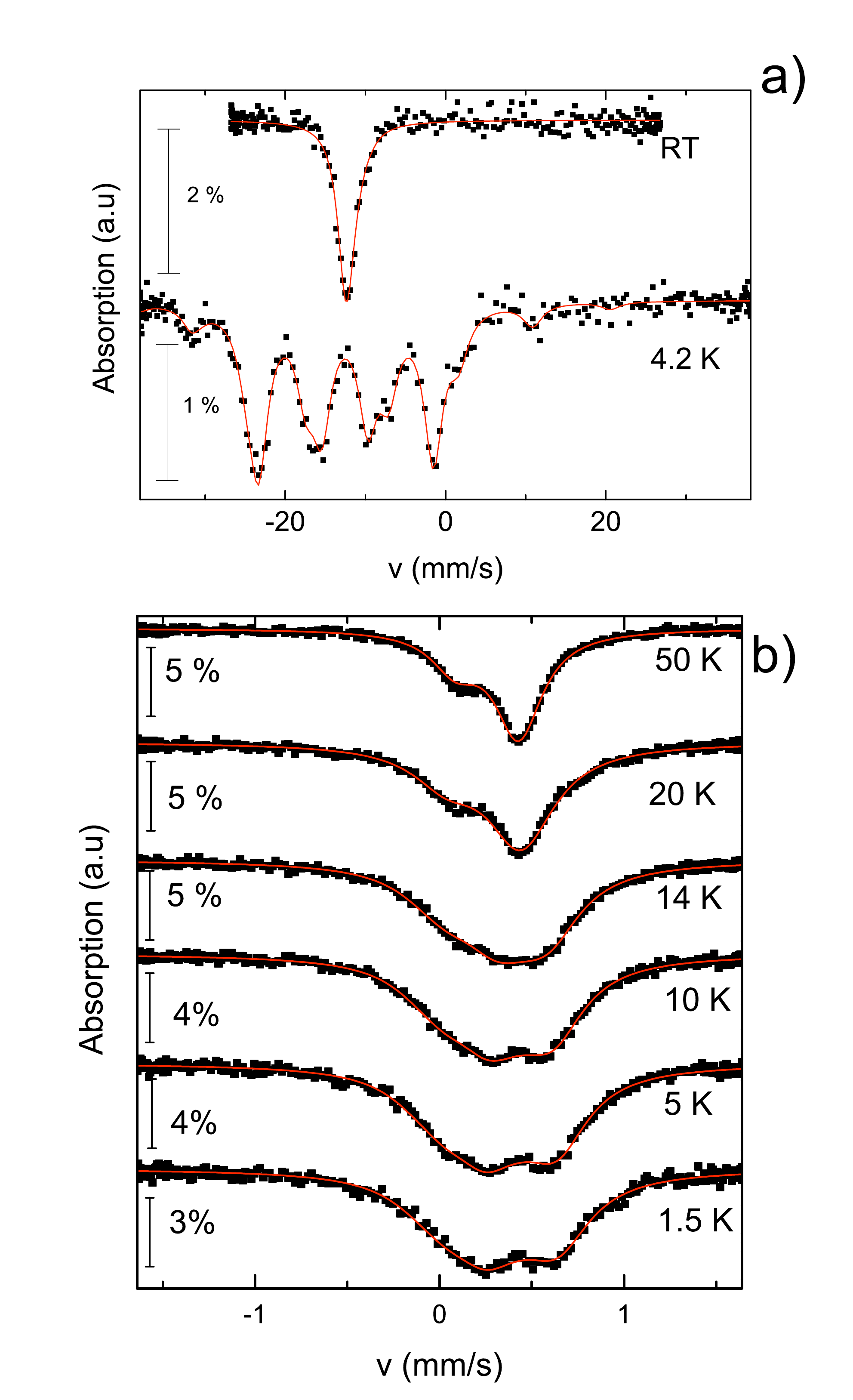} 
\caption{\label{Figure 4.} (a) $^{151}$Eu M\"ossbauer effect spectra for the single crystal mosaic at room temperature and 4.2 K.  (b) $^{57}$Fe M\"ossbauer effect spectra for a EuFe$_2$As$_{1.4}$P$_{0.6}$ single crystal mosaic from 50 K to 1.5 K.}
\end{center}
\end{figure}

$^{151}$Eu Mossbauer spectra were taken at room temperature and at 4.2 K.  With this technique we can not only determine the direction of the Eu moments with respect to the crystallographic axes, but we also can see whether or not there is a Eu valence change due to P doping as proposed by Sun et. al. \cite{lilingsun}, since Eu$^{2+}$ and Eu$^{3+}$ are easily distinguished by their different isomer shift values.  All spectra were fitted with the full Hamiltonian model\cite{ruebenbauer}.  The room temperature spectrum in Fig.~\ref{Figure 4.}(a) does not show any evidence for the Eu$^{3+}$ valence, as only an absorption line related to the Eu$^{2+}$ valence was observed, in agreement with recent polycrystalline work\cite{nowik}. The isomer shift value is $\delta=$-11.8(2) mm/s, and the quadrupole splitting $\Delta E_Q=$-3.5(6) mm/s.  At 4.2 K the spectrum is magnetically split due to the magnetic ordering of the Eu moments.  The hyperfine parameters are $\delta=$-11.8(1) mm/s, $\Delta E_Q=$-4.1(8) mm/s, $B_{HF}$=28.4(2) T, $\theta=12(8)$ degrees, values indicating that the Eu$^{2+}$ moments are aligned almost parallel to the crystallographic c-axis.  This is in agreement with P doping or application of an external magnetic field that breaks the weak coupling between AF coupled Eu layers, and reorients the moments along the c-axis, resulting in a FM ordered state \cite{jiang2,1011.4481,nowik}.  All the hyperfine parameters given above essentially are in agreement with those found in polycrystalline samples of EuFe$_2$As$_{1-x}$P$_{x}$ by Nowik \textit{et. al.} \cite{nowik}.

\begin{figure}[t]
\begin{center}
\includegraphics[angle=0,width=8.0cm]{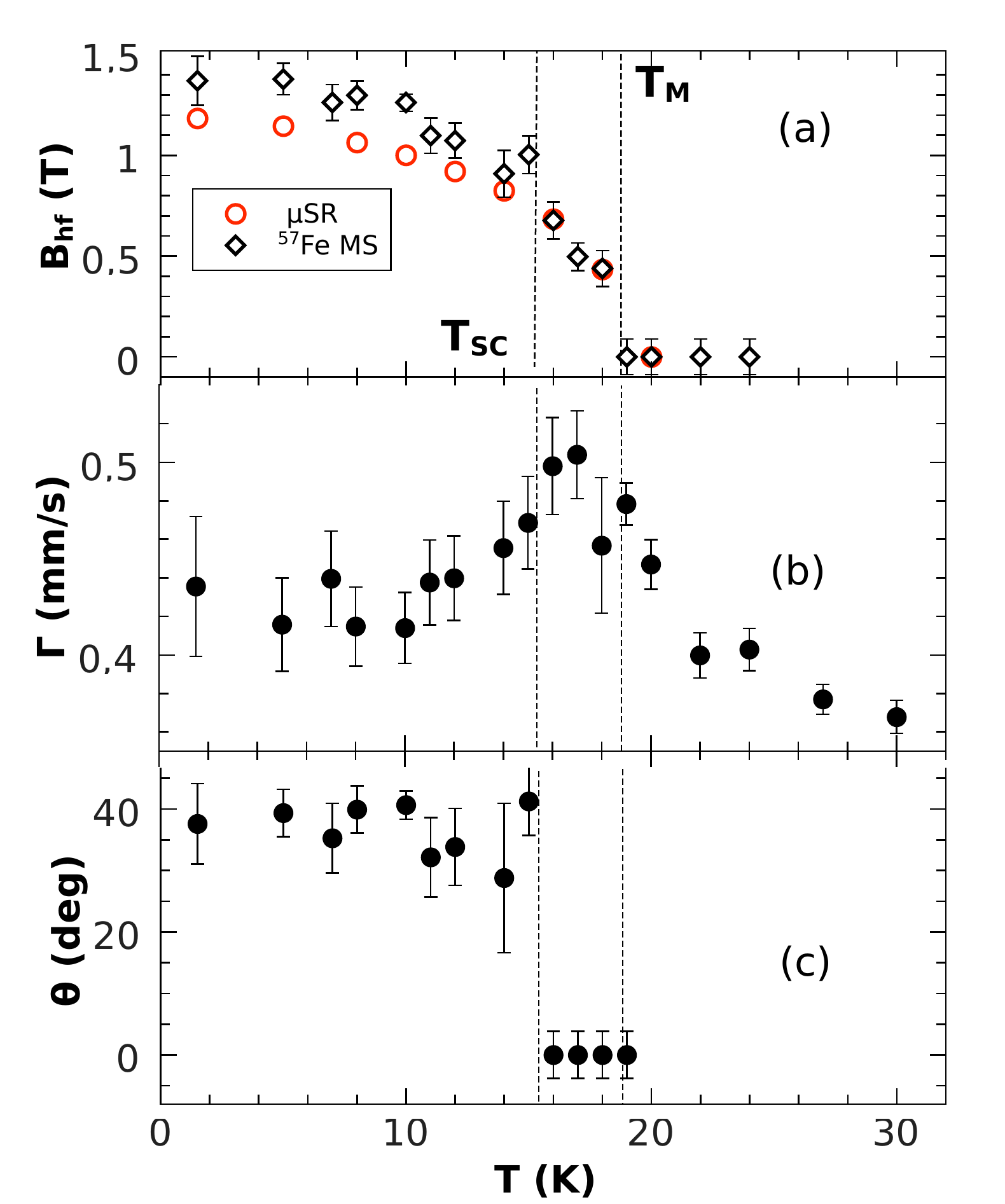} 
\caption{\label{Figure 5.} Magnetic hyperfine parameters extracted from $^{57}$Fe M\"ossbauer Spectra.  The magnetic hyperfine field is compared to the equivalent hyperfine field at the muon site obtained from zero field $\mu$SR; below 15 K the two hyperfine field values deviate. The linewidth shows a maximum close to 15 K, contrary to the muon relaxation rate that peaks close to 18 K. The angle between the hyperfine field and V$_{ZZ}$ shows a jump at 15 K.}
\end{center}
\end{figure}

$^{57}$Fe M\"ossbauer spectra were taken at room temperature and in the range 2 $-$ 50 K; some of these spectra are shown in Fig.~\ref{Figure 4.} (b). The asymmetric shape of the spectrum is due to the orientation of the electric field gradient principal axis $V_{ZZ}$ with respect to $\gamma$ rays.  The intensity ratio of the two quadrupole lines is given by $\frac{I_{\frac{3}{2}}}{I_{\frac{1}{2}}}=\frac{1+\cos^2{\theta}}{\frac{2}{3}+\sin^2{\theta}}$, with $\theta$ being the angle between $V_{ZZ}$ and the $\gamma$-ray direction.  Fitting of the spectrum results in $\theta=$15(2) degrees.  Since $V_{ZZ}$ is parallel to the c-axis \cite{mariella}, as have been found for similar single crystalline compounds, we have to conclude that some of the small single crystal platelets are slightly misoriented.  As the temperature approaches 20 K the linewidth increases continuously.  The fits of these data were performed taking constant $\Delta E_Q$=0.40(1) mm/s below 20 K; this is expected for metallic systems at low temperatures in the absence of a structural phase transition.  For example, the increase in $\Delta E_Q$ observed by Alzamora \textit{et. al.} for CaFe$_2$As$_2$ single crystals \cite{mariella} between 20 K and 4.2 K is only 0.005(5) mm/s.  Below 18 K the shape of the spectrum exhibits a dramatic change, indicating the presence of a transferred magnetic hyperfine field coming from ordered Eu atoms.  The spectra have been fitted with one component only.  This is in sharp contrast to the finding of Nowik \textit{et. al.} \cite{nowik} who measured $^{57}$Fe spectra of polycrystalline EuFe$_2$As$_{1-x}$P$_x$, where two Fe components clearly have been found.  The behavior of the hyperfine field, linewidth, and the angle between the hyperfine field and the c-axis are shown in Fig.~\ref{Figure 5.}. The isomer shift value is $\delta=$ 0.36(1) mm/s, with negligible variation in temperature. 

We can see in Fig.~\ref{Figure 5.} that for $T_{SC}\leq T\leq T_C$, both the transferred magnetic hyperfine fields at the muon site and the $^{57}$Fe nuclei are the same, using the muon gyromagnetic ratio $\tilde{\gamma}=135.54$ MHz/T.  This indicates that the muons are likely sitting within the FeAs(P) layers on sites which are magnetically equivalent to the Fe lattice sites.  Both the muon and Fe nuclei are only sensing the ordering of the Eu magnetic moments occurring below $T_C$ and there seems to be no contribution coming from a Fe magnetic moment.  However, for $T \leq T_{SC}$ we find $B_{hf}(^{57}$Fe$)>B_{hf}(\mu)$ (refer to Fig.~\ref{Figure 5.}(a)) and at the same time an abrupt change in $\theta$ (angle between direction of $B_{hf}$ and c-axis) from $\theta\approx0^o$ to $\theta\approx 40^o$ (Fig. \ref{Figure 5.}(c)).  Also the linewidth $\Gamma$ has a maximum at $T_{SC}$ (see Fig.~\ref{Figure 5.}(b)) and not at $T_C$ as expected.  

All these facts could be explained as follows: the fluctuation rate of the small Fe moments, having a magnitude of the order of $\mu_{Fe}\approx0.07$ $\mu_B$ \cite{ahmed}, is slowing down in the temperature region from 50 K to $T_{SC}$  resulting in an increase in the $^{57}$Fe linewidth.  For $T\leq T_{SC}$  the magnetic moments ``freeze'' within the ab-plane resulting in an increase in the magnetic field at the $^{57}$Fe nucleus (a magnetic moment of $\mu_{Fe}=0.07$ $\mu_B$ corresponds to a magnetic hyperfine field $B_{Fe}=1$ T \cite{dubiel}) and a canting of $\vec{B}_{hf}'=\vec{B}_{hf}+\vec{B}_{Fe}$, with $\vec{B}_{hf}$ and $\vec{B}_{Fe}$ being parallel and perpendicular to the c-axis, respectively.  We are not aware that such a change in the dynamics of the Fe magnetic moment at the onset of superconductivity has been seen in pncitides before.  It is not clear, however, whether this change in the Fe moment dynamics is a necessary condition for forming the superconducting state, or if it is resulting from the formation of the superconducting state.  Independent of this, we have strong evidence that the coexistence between ferromagnetism and superconductivity is indeed present in this pnictide and not an artifact caused by phase separation, as claimed by Jeevan \textit{et. al.} \cite{1011.4481}. This just follows from the two facts: (i) hyperfine parameters from $\mu$SR, $^{57}$Fe and $^{151}$Eu probes clearly show the presence of only one, namely, a magnetic phase and (ii) the onset of superconductivity is reflected in the hyperfine paramenters of this magnetic phase.  There is clearly an interplay between the two phenomena which could be explained by a coupling between localized and itinerant Fe $3d$ electrons \cite{yang}.

{\bf Acknowledgement:\/}  This work has been supported by the US NSF under the Materials World Network (MWN: DMR-0502706 and 0806846) and the Partnership for International Research and education (PIRE: OISE-0968226) programs at Columbia, by Canadian NSERC and CIFAR at McMaster, and by CIAM (CNPq-NSF), CNPq and Faperj at CBPF in Rio de Janeiro, Brazil and NSFC and MOST of China: 973 project 2011CB605900 at IOP in Beijing.  H. Micklitz acknowledges a visitor fellowship of CAPES to work at CBPF.\\

\vfill \eject
\end{document}